# Non-iterative Methods in Inhomogeneous Background Inverse Scattering Imaging Problem Assisted by Swin Transformer Network

Naike Du, Tiantian Yin, Jing Wang, Rencheng Song, Kuiwen Xu, Bingyuan Liang, Sheng Sun and Xiuzhu Ye, *Senior Member, IEEE*

*Abstract*—A deep learning-assisted inversion method is proposed to solve the inhomogeneous background imaging problem. Three non-iterative methods, namely the distorted-Born (DB) major current coefficients method, the DB modified Born approximation method, and the DB connection method, are introduced to address the inhomogeneous background inverse scattering problem. These methods retain the multiple scattering information by utilizing the major current obtained through singular value decomposition of the Green's function and the scattered field, without resourcing to optimization techniques. As a result, the proposed methods offer improved reconstruction resolution and accuracy for unknown objects embedded in inhomogeneous backgrounds, surpassing the backpropagation scheme (BPS) and Born approximation (BA) method that disregard the multiple scattering effect. To further enhance the resolution and accuracy of the reconstruction, a Shifted-Window (Swin) transformer network is employed for capturing super-resolution information in the images. The attention mechanism incorporated in the shifted window facilitates global interactions between objects, thereby enhancing the performance of the inhomogeneous background imaging algorithm while reducing computational complexity. Moreover, an adaptive training method is proposed to enhance the generalization ability of the network. The effectiveness of the proposed methods is demonstrated through both synthetic data and experimental data. Notably, super-resolution imaging is achieved with quasi real-time speed, indicating promising application potential for the proposed algorithms.

## I. Introduction

INHOMOGENEOUS background imaging, i.e., to detect the objects hidden behind the obstacle that cannot be accessed directly, is of great importance in many application scenarios, such as biomedical imaging, through wall imaging, and non-destructive evaluation [1][2]. The inhomogeneous background imaging problem can be solved by radar-based methods, which usually require wide frequency bandwidth for high resolution [4]. Such as in [5], a thorough analysis is presented to incorporate all the reflection and transmission effects induced by the wall. And in [3], the challenge of reconstructing the dielectric structure within layered media is effectively tackled. However, in radar-based methods the reconstructed images are qualitative ones which cannot tell the material of the hidden objects. Moreover, the Born approximation is taken in these methods, which neglects the multiple scattering effect between the inhomogeneous background and the scatterer and thus the super-resolution information is totally lost.

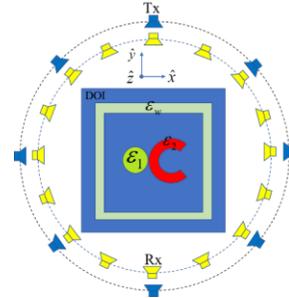

**Fig. 1.** Experimental configuration.

The inverse scattering method enables quantitative reconstruction of the constitutive parameters of scatterers with super resolution (if the minimum resolvable distance of two objects is smaller than half wavelength, it will be defined as super-resolution), due to the conservation of the multiple scattering information [6]. However, the problem is highly nonlinear and is ill-posed as the collected scattered field is far less than the number of unknowns. Typically, it is addressed by minimizing a cost function constructed from the mismatch between the measured scattered field and the calculated counterpart. Nonlinear methods, such as the modified gradient method (MGM) [7], the distorted-Born iterative method (DBIM) [8], the contrast source inversion algorithm (CSI) [9], and the subspace-based optimization method (SOM) [10],

This work is supported by the Natural National Science Foundation of China (NSFC) (No. 62001474, No. 61971036) and the Fundamental Research Funds for the Central Universities.

Naike Du, Jing Wang, and Xiuzhu Ye are with the School of Information and Electronics Engineering, Beijing Institute of Technology, Beijing, China. (Corresponding author email: xiuzhuye@outlook.com)

Tiantian Yin is with the Department of Electrical and Computer Engineering, National University of Singapore, Singapore.

Rencheng Song is with School of Instrument Science and Opto-electronics Engineering, Hefei University of Technology, Hefei, China.

KuiwenXu is with Hangzhou Dianzi University, Hangzhou, China.

Bingyuan Liang is with Institute of Telecommunication and Navigation Satellites China Academy of Space Technology. Beijing, China.

Sheng Sun is with the School of Electronic Science and Engineering, University of Electronic Science and Technology of China, Chengdu, China.

2effectively capture the multiple scattering effect and have proven to be successful in achieving high resolution quantitative imaging.

Due to the substantial number of unknowns involved in pixel-based inversion algorithms, deterministic optimization is commonly employed. However, this approach can be time-consuming due to the need for gradient calculations at each iteration. Consequently, the inverse scattering method is seldomly applied in real-time imaging scenarios. On the other hand, non-iterative inversion algorithms such as the backpropagation scheme (BPS) or Born approximation algorithm offer faster computation. Nevertheless, they suffer from low resolution since they neglect the multiple scattering effect, which plays a crucial role in super-resolution information. In [11], three non-iterative inversion algorithms were proposed for reconstructing scatterers in free space background, namely the major current coefficients method, the modified Born approximation method, and the connection method. In the meantime, they are able to achieve high-resolution reconstruction, as the multiple scattering information is retained through the reconstructed major current.

In inhomogeneous background imaging scenarios, addressing the multiple scattering effect between the background and the unknown scatterer is a key challenge. Therefore, nonlinear inverse scattering approaches are commonly employed to tackle the inhomogeneous background imaging problem. In [12], the multi-layer Green's function is utilized to model scatterers embedded between two walls, with the subspace-based optimization method (SOM) employed as the inversion algorithm. [13] adopts the finite element method to model the wall, while utilizing the contrast source inversion (CSI) method. Additionally, [14] introduces the concept of the separable obstacle problem, where the object does not overlap with the wall. In this case, the wall is treated as a known object rather than part of the background, effectively reducing the nonlinearity of the inhomogeneous background imaging problem. The proposed method demonstrates its effectiveness in reconstructing objects adjacent to the wall.

Recently, the effectiveness of machine learning-based inversion methods in generating high-resolution reconstructions at real-time computational speed has been demonstrated. For instance, in [15] and [16], the backpropagation scheme (BPS) is employed to linearly reconstruct a coarse image of the relative permittivity. This coarse image is then input into a trained U-net convolutional neural network (CNN), which outputs the reconstructed high-resolution image. Similarly, in [17], machine learning techniques are utilized to address the computationally demanding gradient calculations involved in the optimization process. In [18], a cascaded CNN architecture is employed to retrieve the induced current. Moreover, in [19], the application of generative adversarial networks (GANs) eliminates the directional reconstruction challenge associated with anisotropic scatterer inverse scattering. Notably, in [20], machine learning is applied to solve the inhomogeneous background imaging problem. A linear method known as distorted-Born backpropagation scheme (DB-BPS) is proposed to retrieve a coarse image of the concealed object. This coarse image is subsequently fed into a GAN framework to achieve quasi-real-time high-resolution imaging. These studies collectively highlight the advancements and potential of machine learning approaches in addressing inverse scattering challenges.

In this article, we present three non-iterative methods that aim to solve the inhomogeneous background imaging problem with high resolution by effectively preserving the multiple scattering effect without resorting to optimization. Additionally, we introduce a Swin transformer network with an attention mechanism to enhance resolution specifically in scenarios where the contrast of the scatterer is high. The conventional CNN is the backbone networks for a variety of vision tasks and the convolution operation is a "local" operation bounded to a small neighborhood of an image. Compared to the conventional CNN, Swin Transformers uses self-attention, a "global" operation, which draws information from the whole image. The Swin Transformer model can conveniently leverage advanced techniques for dense prediction such as feature pyramid networks or U-Net. These merits make Swin Transformer suitable as a general-purpose backbone for various vision tasks [21-23]. The contributions of our proposed methods can be summarized as follows:

1) Three non-iterative methods, namely the distorted Born (DB) major current coefficient methods, the DB modified Born approximation method, and the DB connection method, are proposed to address the inhomogeneous background imaging problem. These methods involve the analytical retrieval of the major current induced by the unknown scatterer within an inhomogeneous background. By incorporating the multiple scattering information through the utilization of the major current, the scattered field of the scatterer can be accurately calculated, as opposed to being completely neglected as in traditional non-iterative methods. Comparative analysis with existing non-iterative methods, such as DB-BPS [20] and distorted-Born approximation (DBA) method, demonstrates that the proposed methods exhibit superior resolution and higher accuracy. Additionally, a comparative evaluation among the three proposed methods reveals that the DB-MBA method achieves the best performance.

2) To enhance the imaging quality particularly in scenarios with high contrast scatterers, we propose the utilization of a Swin transformer network. Firstly, the Swin transformer network offers the advantage of effectively modeling long-range dependencies through the implementation of a shifted window scheme. This enables the algorithm to capture global interactions between objects, thereby improving the performance of the inhomogeneous background algorithm. Secondly, the Swin transformer exhibits efficient processing capabilities for large-sized images due to its local attention mechanism. The introduction of the shifted window scheme significantly reduces the computational complexity involved in the process. Lastly, in order to optimize the training process, we adopt an adaptive distribution approach for the training dataset based on the mean square errors (MSE) of the coarse images obtained from the non-iterative methods. Consequently, examples with lower MSE receive less emphasis during training, while those with higher MSE are given greater attention. This adaptive distribution strategy enhances the generalization ability of the proposed network, leading to improved overall performance.

3) The effectiveness of the proposed method is evaluated through validation using both synthetic and measured data, demonstrating its real-time super-resolution capability in inhomogeneous background imaging. In order to assess the generalization ability of the network, the training process utilizes the MNIST dataset, while the "Austria profile" and additional experimental examples beyond the scope of the MNIST dataset are employed for rigorous generalization testing purposes. This comprehensive validation process ensures the reliability and robustness of the proposed method in various scenarios, highlighting its potential for practical application in real-world imaging scenarios.

## II. FORWARD PROBLEM

In this article, a 2D transverse magnetic (TM) problem is considered. As depicted in Fig. 1, inside the domain of interests (DOI), the unknown objects (relative permittivity is $\varepsilon_1$ and $\varepsilon_2$ respectively) are surrounded by inhomogeneous background with relative permittivity $\varepsilon_w$. The shape and material of the inhomogeneous background are known *apriori*. The DOI is illuminated by $N_i$ plane waves evenly distributed around a circle. Meanwhile, $N_r$ receiving antennas are evenly distributed around a circle to record the scattered field.

The forward problem is described by the method of moments (MOM). The pulse function is used as the basis function and the delta function is used as the testing function. The DOI $D$ is discretized into $N$ square cells with centers located at $r_1, r_2, ..., r_N$. Within $D$, after discretization, the vectors representing the total field, the incident field, the contrast and the induced current are given by $\bar{E}_n^{\text{tot}} = \bar{E}^{\text{tot}}(r_n)$, $\bar{E}_n^{\text{inc}} = \bar{E}^{\text{inc}}(r_n)$, $\bar{\xi}_n = -i\omega\varepsilon_0[\bar{\varepsilon}_r(r_n) - 1]$, $\bar{J}_n = \bar{J}(r_n)$ respectively, where $\bar{\varepsilon}_r(r_n)$ is the relative permittivity at $r_n$. It satisfies the following discretized Lippmann-Schwinger equation:

$$\bar{E}^{\text{tot}} = \bar{E}^{\text{inc}} + \bar{\bar{G}}_d \cdot \bar{\bar{\xi}} \cdot \bar{E}^{\text{tot}}, \quad (1)$$

where $\bar{\bar{G}}_d$ is the discretization matrix of the 2-D free space Green's function in domain $D$. The equivalent radius of each cell in DOI is donated as $a = \sqrt{S/\pi}$, where $S$ is the area of the discretized square cell. Then, the elements in $\bar{\bar{G}}_d$ are defined as,

$$\bar{\bar{G}}_d(n,n') = \frac{-\eta\pi a}{2} J_1(k_0 a) H_0^{(1)}(k_0 |\bar{r}_n - \bar{r}_{n'}|), \quad (2)$$

for $n \neq n'$, and

$$\bar{\bar{G}}_d(n,n') = \frac{-\eta\pi a}{2} H_1^{(1)}(k_0 a) - \frac{i\eta}{k_0}, \quad (3)$$

for $n = n'$, where $k_0$ is the wavenumber in free space and $\eta$ is intrinsic impedance of free space. And the induced current is expressed as:

$$\bar{J} = \bar{\bar{E}}^{\text{tot}} \cdot \bar{\xi} = \bar{\bar{\xi}} \cdot \bar{E}^{\text{tot}}. \quad (4)$$

Here $\bar{E}^{\text{tot}}$, $\bar{\xi}$ are column vectors of total electric field and normalized contrast, $\bar{\bar{E}}^{\text{tot}}$, $\bar{\bar{\xi}}$ are the matrix with vectors arranged in the diagonal. Combining with (1), the expression of the induced current is then changed as:

$$\bar{J} = \bar{\bar{\xi}} \cdot (\bar{E}^{\text{inc}} + \bar{\bar{G}}_d \cdot \bar{J}). \quad (5)$$

The scattered field at the receiving antenna is:

$$\bar{E}^{\text{sca}} = \bar{\bar{G}}_s \cdot \bar{J}, \quad (6)$$

where $\bar{\bar{G}}_s$ is the mapping of the induced current to the scattered field on the receiving antenna, and $\bar{r}_q^s$ is the vector from transmitting antenna to receiving antenna:

$$\bar{\bar{G}}_s(q,n) = \frac{-\eta\pi a}{2} J_1(k_0 a) H_0^{(1)}(k_0 |\bar{r}_q^s - \bar{r}_n|) \quad (7)$$

Then we derive the governing equations for the inhomogeneous background scenario. The contrast of the background and target objects are denoted as $\bar{\bar{\xi}}^{bac}$ and $\bar{\bar{\xi}}^{obj}$ respectively. Unlike the case of homogeneous background where the free space Green's functions $\bar{\bar{G}}_s$ and $\bar{\bar{G}}_d$ are utilized, we extend our approach to incorporate the influence of the background into the inhomogeneous background Green's functions, namely $\bar{\bar{G}}_{bs}$ and $\bar{\bar{G}}_{bd}$ [20][24],

$$\bar{\bar{G}}_{bs} = \bar{\bar{G}}_s \cdot (\bar{\bar{I}} - \bar{\bar{\xi}}^{bac} \cdot \bar{\bar{G}}_d)^{-1}, \quad (8)$$

$$\bar{\bar{G}}_{bd} = \bar{\bar{G}}_d \cdot (\bar{\bar{I}} - \bar{\bar{\xi}}^{bac} \cdot \bar{\bar{G}}_d)^{-1}. \quad (9)$$

where $\bar{\bar{G}}_{bs}$ denotes the mapping from the induced current on scatterers embedded in inhomogeneous background to the scattered field on receivers, and $\bar{\bar{G}}_{bd}$ denotes the mapping inside the domain of interest.

The total electric field due to the background is given by,

$$\bar{E}^{bac} = (\bar{\bar{I}} - \bar{\bar{G}}_d \cdot \bar{\bar{\xi}}^{bac})^{-1} \cdot \bar{E}^{inc}. \quad (10)$$

It can also be understood as the secondary incident field in presence of the inhomogeneous background. The total field $\bar{E}^{tot}$ can thus be expressed as summation of the secondary incident field and the scattered field produced by the object in presence of the inhomogeneous background,

$$\bar{E}^{tot} = \bar{E}^{bac} + \bar{\bar{G}}_{bd} \cdot \bar{J}^{obj}, \quad (11)$$

which is the state equation for inhomogeneous background. And here

$$\bar{J}^{obj} = \bar{\bar{\xi}}^{obj} \cdot \bar{E}^{tot}, \quad (12)$$

denotes the induced current produced by the object in presence of the inhomogeneous background.

The scattered field of the object on the receiver can be calculated by extracting the scattered field due to the

background from the total scattered field, and the data equation for inhomogeneous background is given as

$$\bar{E}^s = \bar{E}^{sca} - \bar{\bar{G}}_s \cdot \bar{\bar{\xi}}^{bac} \cdot \bar{E}^{bac} = \bar{\bar{G}}_{bs} \cdot \bar{J}^{obj}. \quad (13)$$

## III. INVERSION ALGORITHM

The purpose of the inverse scattering problem is to reconstruct the distribution of relative permittivity in the DOI using the measured scattered field. Three non-iterative methods are proposed to solve the inhomogeneous background imaging problem. When the contrast of the object is high, a Swin transformer network is proposed to further enhance the resolution of the image.

### A. Distorted-Born (DB) Major Current Coefficients Method

The right singular vectors of $\bar{\bar{G}}_{bs}$ serve as basis for $\bar{J}^{obj}$. The singular value decomposition of matrix $\bar{\bar{G}}_{bs}$ is written as

$$\bar{\bar{G}}_{bs} = \bar{\bar{U}} \cdot \bar{\bar{\Sigma}} \cdot \bar{\bar{V}}^H, \quad (14)$$

and the induced current produced by the object in presence of the inhomogeneous background is represented as:

$$\bar{J}^{obj} = \bar{\bar{V}} \cdot \bar{\alpha}, \quad (15)$$

where $\bar{\alpha}$ is a column vector of induced current coefficients. And the elements of $\bar{\alpha}$ are given by

$$\alpha_j = \frac{\bar{u}_j^H \cdot \bar{E}^s}{\sigma_j}, \quad (16)$$

where $\sigma_j$ is the $j$-th singular value of $\bar{\bar{G}}_{bs}$, $\bar{u}_j^H$ represents the left eigenvector corresponding to the $j$-th singular value, and H is the Hermitian of the matrix.

$\bar{\bar{G}}_{bs}$ is a compact matrix which maps the induced contrast current (due to the profile $\bar{\bar{\xi}}^{obj}$) to the scattered field on the receiver. There is one portion of current that produces the scattered field on receiver, whereas another portion produces zero or negligible scattered field. We arrange the singular values in a descending order such that $\sigma_1 > \sigma_2 > ... > \sigma_{L_0} > \sigma_{L_0+1} = \sigma_{L_0+2} = ... = \sigma_N = 0$.

We can mathematically retrieve one portion of the radiating current using the largest $L$ leading singular values, the major induced current is then defined as:

$$\bar{J}^+ = \bar{\bar{V}}^+ \cdot \bar{\alpha}^+ \quad (17)$$

where + means the major part using the largest $L$ singular values. It should be noted that thin SVD [6] is used in this article, rather than the full SVD, allowing for substantial savings in terms of computational resources and processing time.

Combining (17) and (12), the expression for the main current coefficient can be obtained by

$$\bar{\alpha}^+ = \bar{\bar{V}}^{+H} \cdot \bar{\bar{E}}^t \cdot \bar{\xi}^{obj}, \quad (18)$$

where $\bar{E}^t = \bar{E}^{bac} + \bar{\bar{G}}_{bd} \cdot \bar{J}^+$ denotes the approximated total field got by the major current.

By adding a regularization term, the loss function can be constructed as,

$$f(\bar{\xi}^{obj}) = \sum_{p=1}^{N_i} \left\| \bar{\bar{V}}^{+H} \cdot \bar{\bar{E}}_p^t \cdot \bar{\xi}^{obj} - \bar{\alpha}_p^+ \right\|^2 + \beta \left\| \bar{\xi}^{obj} \right\|^2 \quad (19)$$

where $\beta$ is the Tikhonov regularization parameter. The minimum of $f(\bar{\xi}^{obj})$ requires the derivative with respect to $\bar{\xi}^{obj}$ to be 0, which can obtain the analytical solution of $\bar{\xi}^{obj}$,

32

$$\cdot \left[ \sum_{p=1}^{N_i} (\bar{\bar{V}}^{+H} \cdot \bar{\bar{E}}_p^t)^H \cdot \bar{\alpha}_p^+ \right] \quad (20)$$

We call this method as the distorted-Born (DB) major current coefficients method (DB-MCC).

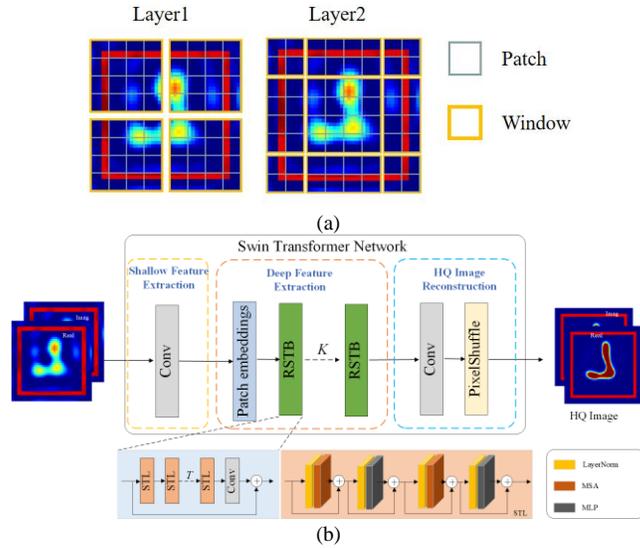

**Fig. 2.** (a)Architecture of shifted window scheme (b) Architecture of SwinIR

## B. Distorted-Born (DB) Modified Born Approximation Method

Under the assumption that the scatterers are weak ones, that is, when the permittivity of scatterers only differs slightly from that of the free space, the inverse scattering problem can be solved by the first order Born approximation. In the case of inhomogeneous background, the total field $\bar{E}^{tot}$ can be approximated by the secondary incident field $\bar{E}^{bac}$ in presence of the inhomogeneous background, by neglecting the scattered field due to the scatterer. This can be understood as the distorted Born approximation (DBA) method by totally neglecting the scattered field produced by the scatterer.

The cost function is constructed using the data equation as:

$$f(\bar{\xi}) = \sum_{p=1}^{N_i} \left\| \bar{\bar{G}}_{bs} \cdot \bar{\bar{E}}_p^{bac} \cdot \bar{\xi}^{obj} - \bar{E}_p^s \right\|^2 + \beta \left\| \bar{\xi}^{obj} \right\|^2 \quad (21)$$

and the solution in the least square sense is:

$$\bar{\xi}^{obj} = \left[ \sum_{p=1}^{N_i} (\bar{\bar{G}}_{bs} \cdot \bar{\bar{E}}_p^{bac})^H \cdot \bar{\bar{G}}_{bs} \cdot \bar{\bar{E}}_p^{bac} + \beta \bar{\bar{I}} \right]^{-1}$$
$$\cdot \left[ \sum_{p=1}^{N_i} (\bar{\bar{G}}_{bs} \cdot \bar{\bar{E}}_p^{bac})^H \cdot \bar{E}_p^s \right] \quad (22)$$

The DBA method along with the DB-BPS as in [20] serves as a comparison to the following three newly proposed methods.

The DBA has totally neglected the scattered field contributed by the scatterer, and thus the nonlinearity information of the high contrast scatterer is lost. A modified Born approximation

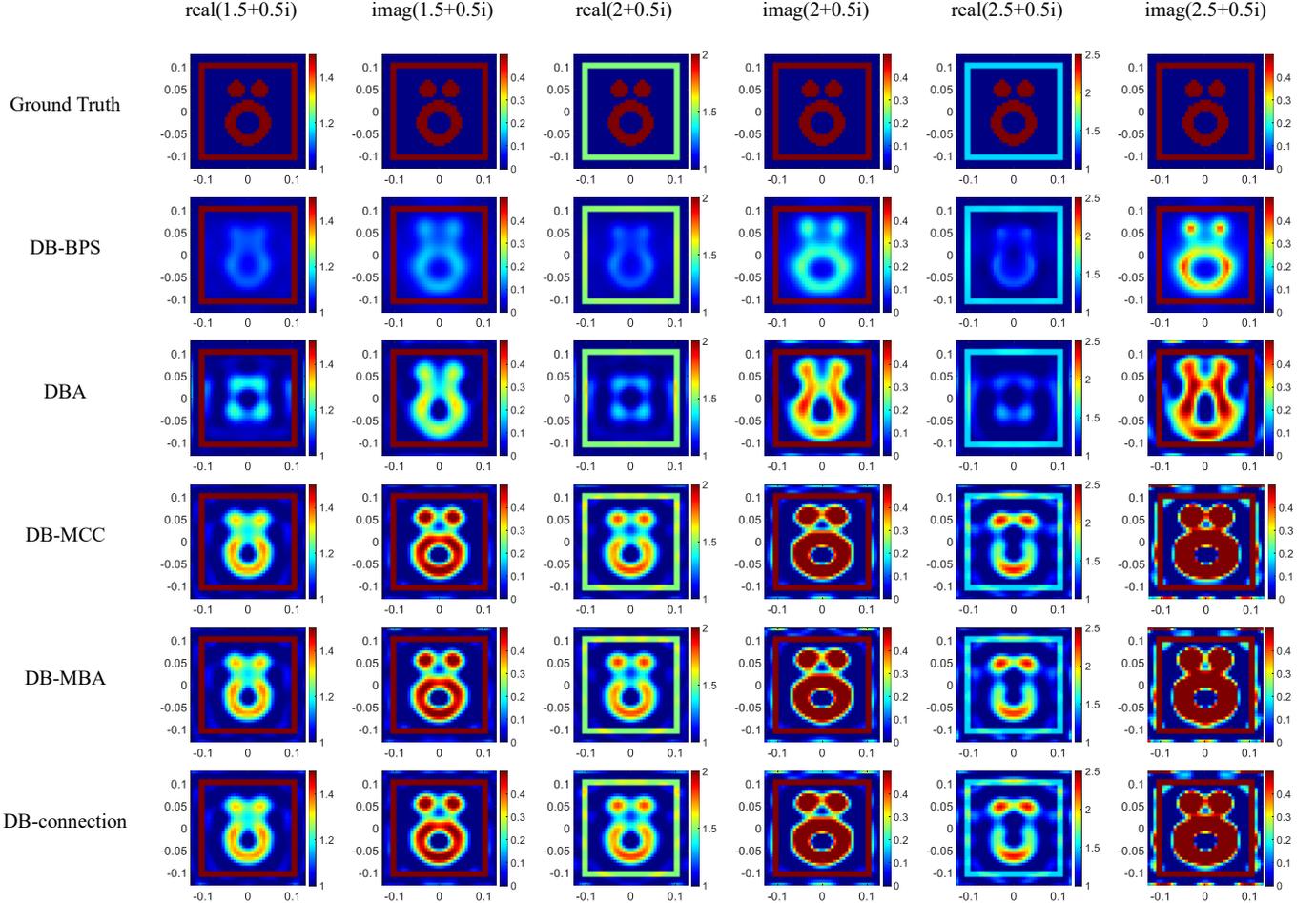

**Fig.3.** Austria profile: Each column represents the relative permittivity distribution reconstructed by different algorithms. Each row represents the reconstruction results of the real and imaginary parts of objects with different contrasts when the inhomogeneous background is 1.5+0.5i.

**Table I**. The errors (MSE and SSIM) for all the examples

| relative permittivity | error | DB-BPS | DBA | DB-MCC | DB-MBA | DB-connection |
|---|---|---|---|---|---|---|
| 1.5+0.5i | MSE | 0.0393 | 0.0339 | 0.0191 | 0.0189 | 0.0193 |
| | SSIM | 0.8531 | 0.8691 | 0.8911 | 0.8920 | 0.8900 |
| 2+0.5i | MSE | 0.0993 | 0.1136 | 0.0754 | 0.0745 | 0.0760 |
| | SSIM | 0.8447 | 0.8505 | 0.8709 | 0.8710 | 0.8683 |
| 2.5+0.5i | MSE | 0.3154 | 0.3594 | 0.2918 | 0.2900 | 0.2935 |
| | SSIM | 0.8586 | 0.8236 | 0.8406 | 0.8412 | 0.8383 |

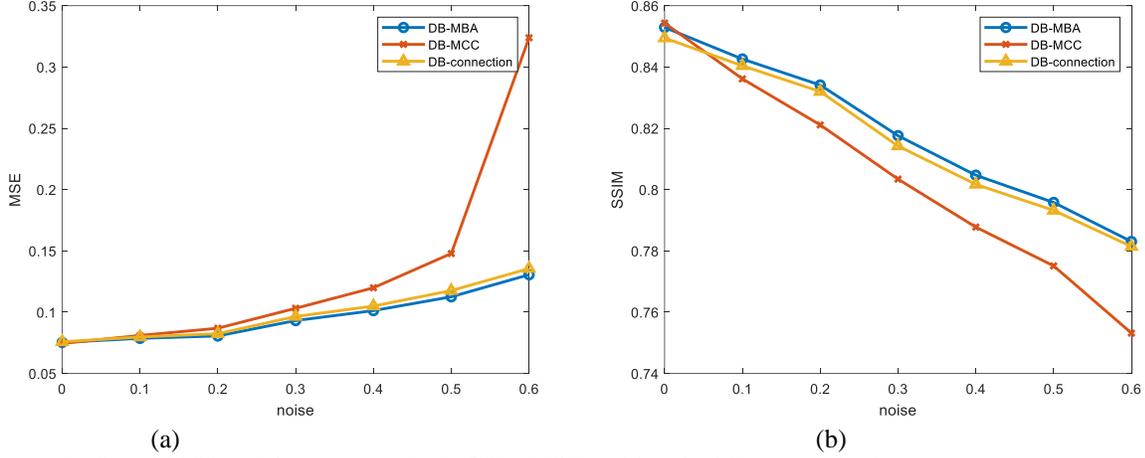
(a) (b)
**Fig.4** average (a) MSE and (b) SSIM of the imaging results for 500 'MNIST' models under different noise levels method for inhomogeneous background, namely the distorted-Born modified Born approximation (DB-MBA) method is proposed here by adding the scattered field generated by the major current to the total field without resourcing to optimization. The cost function is constructed as

$$f(\bar{\xi}^{\mathrm{obj}}) = \sum_{p=1}^{N_i} \left\| \bar{\bar{G}}_{bs} \cdot \bar{\bar{E}}_p^t \cdot \bar{\xi}^{obj} - \bar{E}_p^s \right\|^2 + \beta \left\| \bar{\xi}^{obj} \right\|^2 \quad (23)$$

where $\bar{E}^t = \bar{E}^{bac} + \bar{\bar{G}}_{bd} \cdot \bar{J}^+$ denotes the approximated total field got by the major current. The solution in the least square sense is given by,

$$\bar{\xi}^{obj} = \left[ \sum_{p=1}^{N_i} (\bar{\bar{G}}_{bs} \cdot \bar{\bar{E}}_p^t)^H \cdot \bar{\bar{G}}_{bs} \cdot \bar{\bar{E}}_p^t + \beta \bar{\bar{I}} \right]^{-1} \cdot \left[ \sum_{p=1}^{N_i} (\bar{\bar{G}}_{bs} \cdot \bar{\bar{E}}_p^t)^H \cdot \bar{E}_p^s \right] \quad (24)$$

*C. Distorted-Born (DB) Connection Method*

Multiplying (23) by $\bar{\bar{U}}^{+H}$, the DB-MBA and DB-MCC methods can be connected together, where $\bar{\bar{U}}^{+H}$ is composed by the first $L$ left singular vectors of $\bar{\bar{G}}_{bs}$.

The initial cost function is expressed as,

$$f(\bar{\xi}^{obj}) = \sum_{p=1}^{N_i} \left\| \bar{\bar{U}}^{+H} \cdot (\bar{\bar{G}}_{bs} \cdot \bar{\bar{E}}_p^t \cdot \bar{\xi}^{obj} - \bar{E}_p^s) \right\|^2 + \beta \left\| \bar{\xi}^{obj} \right\|^2 \quad (25)$$

Combined with (16) the cost function is rearranged as,

$$f(\bar{\xi}^{obj}) = \sum_{p=1}^{N_i} \left\| \bar{\bar{\sigma}}^+ \cdot \bar{\bar{V}}^{+H} \cdot \bar{\bar{E}}_p^t \cdot \bar{\xi}^{obj} - \bar{\bar{\sigma}}^+ \cdot \bar{\alpha}_p^+ \right\|^2 + \beta \left\| \bar{\xi}^{obj} \right\|^2 \quad (26)$$

where $\bar{\bar{\sigma}}^+$ is a $L \times L$ diagonal matrix and the *j*th diagonal element of it equals to the *j*th singular value of $\bar{\bar{G}}_{bs}$. The solution to (26) is given by:

$$\bar{\xi}^{obj} = \left[ \sum_{p=1}^{N_i} (\bar{\bar{\sigma}}^+ \cdot \bar{\bar{V}}^{+H} \cdot \bar{\bar{E}}_p^t)^H \cdot \bar{\bar{\sigma}}^+ \cdot \bar{\bar{V}}^{+H} \cdot \bar{\bar{E}}_p^t + \beta \bar{\bar{I}} \right]^{-1} \cdot \left[ \sum_{p=1}^{N_i} (\bar{\bar{\sigma}}^+ \cdot \bar{\bar{V}}^{+H} \cdot \bar{\bar{E}}_p^t)^H \cdot (\bar{\bar{\sigma}}^+ \cdot \bar{\alpha}_p^+) \right] \quad (27)$$

*D. Machine Learning Process*

In this article, the Swin Transformer[22] is adopted to learn and reconstruct the super-resolution information from the coarse images obtained by the non-iterative method.

The Swin Transformer incorporates a shifted window scheme that effectively restricts self-attention calculations to non-overlapping local windows. Simultaneously, the cross-window connections enable the algorithm to capture the global interactions between objects of interest. This hierarchical structure ensures that the computational complexity remains linear with respect to the image size. Specifically, the shifted window mechanism dynamically shifts the window partition across consecutive self-attention layers. As illustrated in Fig. 2 (a), an initial uniform window division is implemented in layer 1, where self-attention is computed within each window. Subsequently, in layer 2, the window partition is shifted, resulting in the creation of new windows. Notably, the self-attention computation within these new windows extends beyond the boundaries of the previous windows in layer 1, establishing connections between them. This cross-window interaction facilitates global information exchange between different regions, thereby enhancing the performance of imaging.

The network employed in this study, as depicted in Fig. 2 (b), comprises three essential modules: the shallow feature extraction module, deep feature extraction module, and high-quality image reconstruction module [22].

The shallow feature module leverages a convolutional layer to effectively extract low-frequency information from the image. This specific design not only facilitates rapid

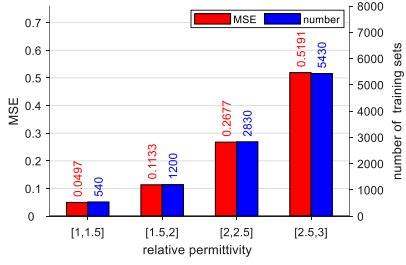

**Fig.5.** The adaptive distribution of the training examples: number is proportional to the average MSE of the input

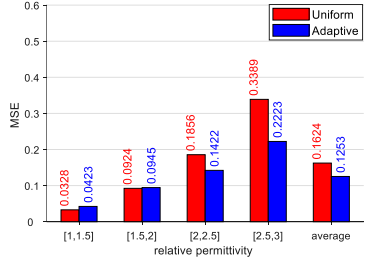

**Fig.6.** The comparison of the MSE for uniform distribution and adaptive distribution of the training examples

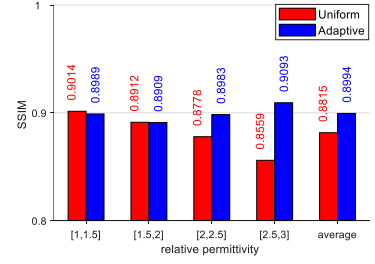

**Fig.7.** The comparison of the SSIM for uniform distribution and adaptive distribution of the training examples

convergence but also ensures the stability of the training process.

The deep feature module plays a crucial role in recovering the lost high-frequency information and is consisted of K residual Swin transformer blocks (RSTB). Initially, the feature map generated by the shallow feature extraction module undergoes a process of division into multiple non-overlapping patch embeddings. In the standard Transformer architecture, for 2D images, a learnable embedding sequence needs to be constructed. This involves dividing the image into numerous non-overlapping patches and transforming the 2D sequence into a 1D sequence through linear projection. Finally, location information is added to the 1D patches, the process of which is referred as patch embeddings. These patch embeddings are then fed into multiple concatenated RSTBs. To maintain the same dimension as the input feature map, numerous non-overlapping patch embeddings are recombined, followed by a convolutional layer. Each RSTB incorporates a residual connection, ensuring the preservation of information flow. Moreover, each RSTB consists of T Swin Transformer Layers (STL). Each STL includes a normalization layer (LayerNorm) followed by a multi-head self-attention (MSA) module, which extracts information from different subspaces. At the end of the MSA, residuals are introduced. To establish full connection between different layers, a LayerNorm followed by a multi-layer perceptron (MLP) is introduced. Overall, the deep feature module employs a hierarchical structure of RSTBs and STLs to capture and restore the lost high-frequency information.

The image reconstruction module consists of convolution and PixelShuffle and the residual is constructed by the difference between low-quality images and high-quality images.

Both the mean squared error (MSE) and Structure Similarity Index Measure (SSIM) are included in the loss function [25-27]:

$$L_{\text{full}} = L_{\text{MSE}} + \chi L_{\text{SSIM}} \tag{28}$$

where $\chi$ is the weighing parameter that represents the proportion of $L_{\text{SSIM}}$ that accounts for the $L_{\text{full}}$.

In summary, the Swin transformer has the following advantages for solving the inhomogeneous background imaging problem:

1) The shifted window scheme in the algorithm facilitates the capturing of global interactions between objects within the image. At the same time, the MSA module utilized in the network allows for the consideration of information from distinct subspaces across various locations of the image concurrently [28]. This approach aligns with the concept employed in the subspace-based optimization method (SOM), where the current is decomposed into different subspaces. As a result, the inhomogeneous background imaging algorithm exhibits enhanced performance, enabling improved reconstruction outcomes.

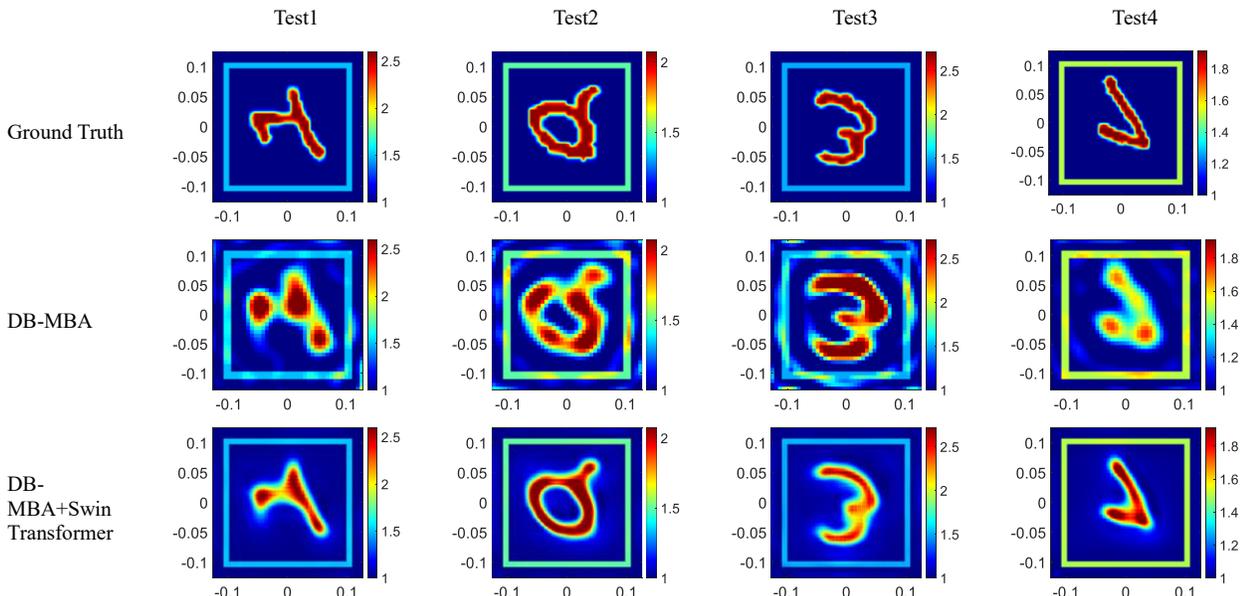

**Fig.8.** Digital objects: reconstructed relative permittivity profiles from scattered fields with 10% noise for DB-MBA and Swin Transformer. The first column shows the ground truth images for four representative tests.

2) The incorporation of the shifted-window approach has effectively reduced the computational complexity within the network, resulting in efficient and optimized modeling. By calculating self-attention within local windows that evenly divide the image in a non-overlapping manner [29][30], the computational burden is significantly alleviated. Assuming an image size of $H \times W$, where each window contains $M \times M$ patches, the computational complexity of both the global multi-head self-attention (MSA) module and the window-based MSA are as follows:

$$\Omega(\text{MSA}) = 4HWC^2 + 2(HW)^2 C \quad (29)$$

$$\Omega(\text{W-MSA}) = 4HWC^2 + 2M^2 HWC \quad (30)$$

The shifted-window method can save a lot of computational resources, given that $M^2$ is significantly lower than $HW$ and in this article $HW = 64 \times M^2$.

3) As the contrast between the target object and the background increases, the error between the coarse image generated by non-iterative method and the ground truth will increase. Therefore, for different contrast distribution intervals, the training difficulties of the network are different. In order to make better use of training resources, this paper introduces the adaptive distribution of the training dataset. The number of trained samples in different contrast intervals is weighted by the average MSE of the coarse images in the range. The adaptive training method can improve the generalization ability of the network. The details of the adaptive training method will be further illustrated in the numerical part.

## IV. NUMERICAL AND EXPERIMENTAL EXAMPLE

In this section, both the synthetic and experimental data are presented to validate the proposed method. In the first four numerical examples, the frequency of the incident wave is 2.4GHz. The DOI of dimension $2\lambda \times 2\lambda$ is discretized into $40 \times 40$ square subunits, the inhomogeneous background is composed by a square obstacle with a length of 21cm and a thickness of 1cm. 12 plane waves evenly distributed around a circle are used to illuminate the DOI one by one. 24 receiving antennas uniformly distributed on a circle with radius 1.13 m are used to collect the scattered field data. The synthetic data is calculated by method of moments, which is contaminated with 10% white Gaussian noise.

*A. First Example: Comparisons of the Non-iterative Methods*

In this example, five non-iterative methods, namely, the DB-BPS, the DBA, the DB-MCC, DB-MBA and the DB-connection method, are compared with each other. The unknown scatterer is "Austria" profile, which is consisted by two discs and one ring. The centers of the two disks and rings are located at $(0.2, 0.4)\lambda$, $(-0.2, 0.4)\lambda$ and $(0, -0.2)\lambda$ respectively. The radius of the discs is $0.15\lambda$. The inner radius of the ring is $0.2\lambda$ and the outside radius is $0.4\lambda$. The relative permittivity of the wall is fixed as $1.5 + 0.5i$. The relative permittivity of the Austria profile changes from 1.5 +0.5i, 2+ 0.5i to 2.5+0.5i. The singular value truncation number $L$ is chosen as 16. The regularization parameter $\beta$ is chosen by the L-curve method [31].

As depicted in Fig. 3, the reconstructed results obtained from the five non-iterative methods are presented and compared. The corresponding estimated errors are listed in Table I. Based on the analysis of the results, the following conclusions can be drawn:

Firstly, the proposed methods, namely DB-MBA, DB-MCC, and DB-connection, demonstrate superior reconstruction capabilities compared to DB-BPS and DBA. This improvement can be attributed to the preservation of the multiple scattered field information of the unknown scatterer. Notably, the sub-wavelength structures of the "Austria" scatterer are accurately reconstructed, achieving super-resolution.

Secondly, among the three proposed methods, DB-MBA exhibits the best overall performance in terms of reconstruction accuracy and resolution. To further improve the conclusion statically, scatterers composed by the 'MNIST' dataset is used. The relative permittivity of the targets is selected randomly within the real part range of (1, 3) and the imaginary part range of (0, 1). The inhomogeneous background is of relative permittivity 1.5+0.5i. We added Gaussian white noise ranging from 0% to 60% to the scattered field data. With a fixed value of $L$=16, we evaluated the imaging results in terms of MSE and SSIM using a predetermined set of 500 'MNIST' models at each noise level. Fig.4 illustrates the outcomes, indicating the consistent advanced performance of the DB-MBA algorithm.

Lastly, it is observed that the reconstruction error increases with the contrast of the scatterer. In cases when the contrast is high, the non-iterative methods may fail to provide the satisfactory reconstruction results. Therefore, a deep learning method is proposed to further enhance the image quality in such scenarios.

*B. Second Example: Digital Objects*

In this example, Swin Transformer is applied to improve the image quality of the non-iterative method. As shown in Fig. 2, the inputs are the reconstructed images got from the DB-MBA, and the outputs are the high-quality reconstructed images with high resolution information.

The scatterers composed by the digital numbers in MNIST data set are used for both training and testing of the Swin Transformer. The relative permittivity of the wall is 1.5. The training set is composed of 10000 samples, while the testing set is composed of 500 samples. The learning rate is set to 0.0002 in the first 30 epoch, and 0.00002 from epoch 30 to epoch 80, and down to 0.000002 in the last 20 epoch. The Swin Transformer is consisted by 6 RSTB layers, each of which contains 2 STL layers. The weighing parameter $\chi$ in the loss function is 0.00025.

Instead of uniformly distributing the number of examples in the range of relative permittivity, the training dataset is adaptively distributed between 1 and 3 according to the mean square errors (MSE) of the coarse reconstructed images for the DB-MBA method. The distributed ratio is as shown in Fig. 5. Therefore, the coarse images with low MSE are less trained while the ones with high MSE can be more trained. The comparison of the uniform and adaptive distributions are as shown in Fig.6 and Fig.7, which reveals that the trained error for higher contrast is greatly improved by the adaptive method.

Some representative ground truth examples are presented in the first row of Fig. 8, the input images got by DB-MBA are shown in the second row, and the outputs of Swin Transformer are as shown in the third row. It can be seen that the resolution of the output images are greatly improved than the input images. The structural similarity (SSIM) index for the whole testing set is calculated as 0.9267.

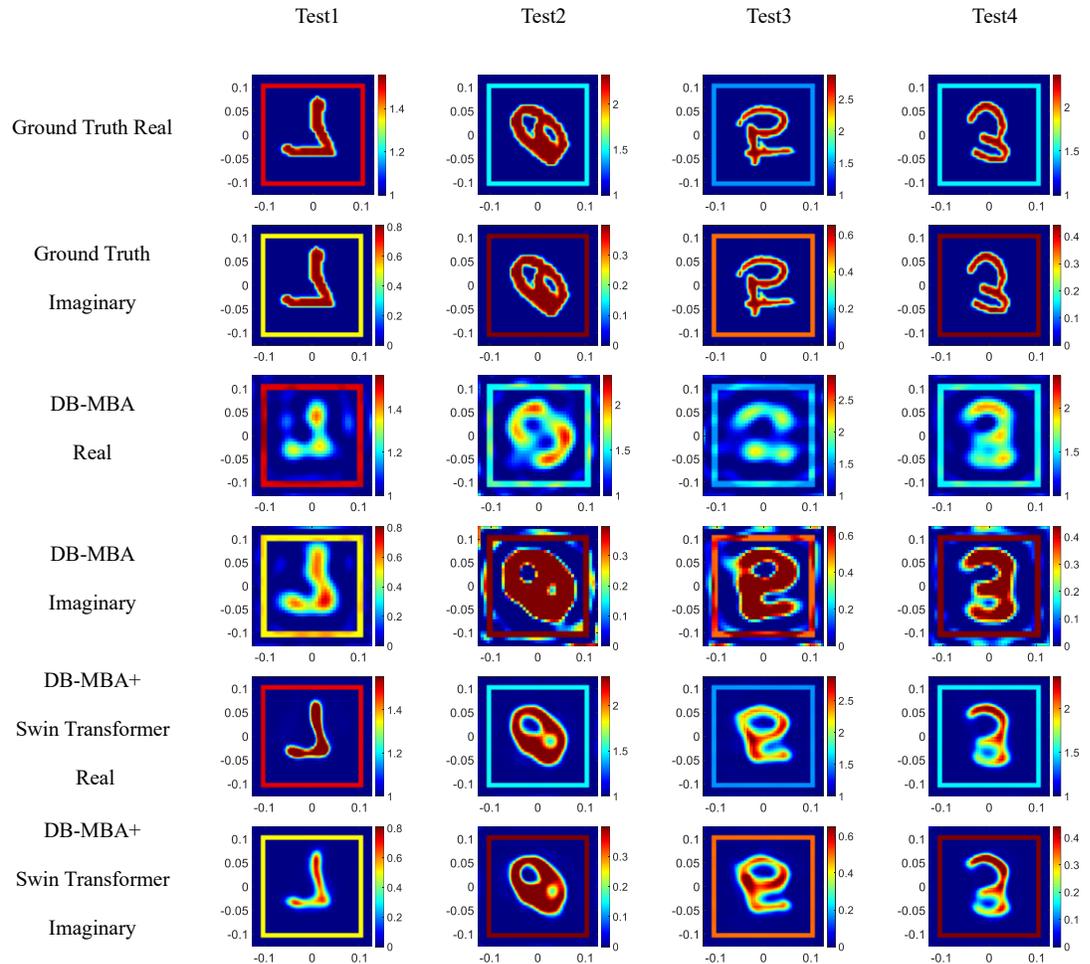

Fig.9. Lossy digital object: the relative permittivity profile is reconstructed from the scattered field with 10% noise, where the real part of the relative permittivity is between 1.5 and 3, and the imaginary part is between 0 and 1.

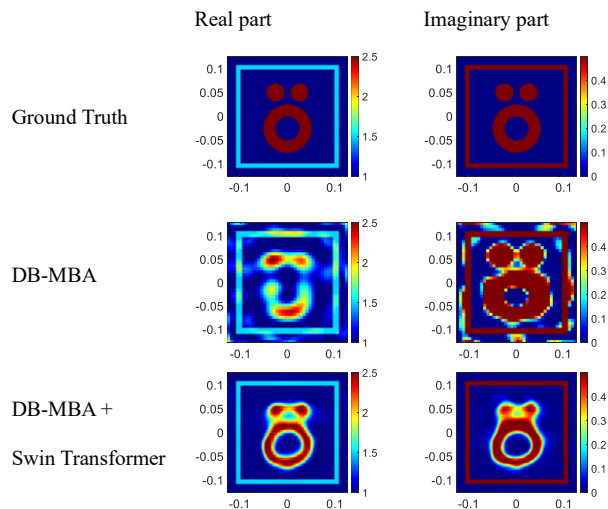

Fig.10. Lossy "Austria" profile: for DB-MBA and Swin Transformer, the relative permittivity profile is reconstructed from the scattered field with 10%

*C. Third Example: Lossy Digital Objects*

In the third example, lossy digital objects are used to test the effectiveness of the proposed method. The number of training examples is adaptively distributed between relative permittivity ranges from 1 to 3 according to the ratio as shown in Fig. 5. And the imaginary part is uniformly distributed from 0 to 1. The training process and hyperparameters of the network are the same as the second example.

In Fig. 9, several testing experiments are presented to show the images of the reconstructed real and imaginary relative permittivity. The SSIM index for the whole testing set is calculated as 0.9153.

*D. Fourth Example: Lossy "Austria" Profile*

In the fourth example, we use the "Austria" profile as the unknown scatterers, the structure of which is as shown in the first row in Fig. 10. The structure is the same as in example one. The relative permittivity of the "Austria" profile and the wall are 2.5+0.5i and 1.5+0.5i respectively. It should be noted that there are no circular scatterers in the training dataset, and this example serves as a test to the generalization ability of the trained Swin Transformer. From the result as shown in the third row of Fig.9, the "Austria" profile is well reconstructed with the subwavelength structure clearly seen. The SSIM for this example is 0.9046.

*E. Fifth Example: Experimental Result*

The experimental data is further tested to verify the proposed method. In the configuration of the through-wall imaging experiment, the operating frequency is set to 2.4 GHz. There are 24 receiving antennas evenly distributed on a circle with diameter 113 cm, 12 of which are used as transmitting antenna as well. The wall is square shaped with side length of 21cm and thickness of 1cm. The relative permittivity of the walls is 2 (Teflon). The unknown scatterers are as shown in the first row in Fig.11, and the relative permittivity for the unknown scatterers is 3 (plexiglass) [32].

From the reconstructed results we can see that, DB-MBA gets a better result than DB-BPS. And by applying the Swin Transformer, the image can be further improved with higher accuracy and resolution. The total time used for reconstruction is only 0.2012 second, which reveals a promising real-time through wall imaging possibility.

**Table II.** Computational time

|  | time/second |
|---|---|
| DB-MBA | 0.1558 |
| Swin Tansformer | 0.0454 |
| total | 0.2012 |

V. CONCLUSION

In this paper, a deep learning-assisted inversion method is proposed to solve the inhomogeneous background inverse scattering problems. Three non-iterative methods based on major current analysis are firstly proposed, which are called DB-MCC, DB-MBA and DB-connection respectively. Then in the high contrast case, the image obtained by the non-iterative algorithm is used as the input of the Swin Transformer network to generate a high-resolution output image.

The contributions of the deep learning-assisted inversion method are as follows: firstly, compared with the traditional DB-BPS and DBA method, the images obtained by the three proposed non-iterative methods achieve higher resolution and better accuracy reconstruction due to the conservation of the multiple scattered field through the major current analysis. Secondly, the Swin transformer network is adopted to further retrieve the super-resolution information of the image. The attention mechanism involved in shifted window enables the algorithm to capture the global interactions between the objects, thus improving the performance of the inhomogeneous background imaging and at same time reducing the computational complexity. Thirdly, both synthetic and experimental data are given to verify the proposed method. Super-resolution imaging is achieved with quasi real-time speed, which reveals a promising application potential of the proposed algorithms.

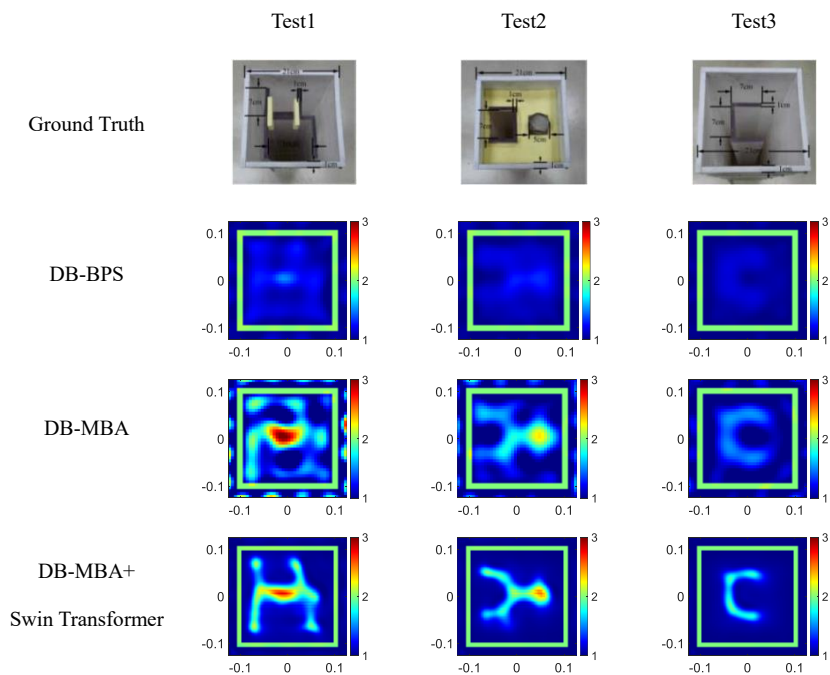

Fig.11. Experimental Result: reconstructed relative permittivity profiles from measured scattered field for DB-MBA and Swin Transformer. The first column shows the ground truth images for three representative tests.

ACKNOWLEDGMENT

This work is supported by the Natural National Science Foundation of China (NSFC) (No. 62001474, No. 61971036) and the Fundamental Research Funds for the Central Universities.

REFERENCES

[1] Q. Huang, L. Qu, B. Wu and G. Fang, "UWB Through-Wall Imaging Based on Compressive Sensing," IEEE Transactions on Geoscience and Remote Sensing, vol. 48, no. 3, pp. 1408-1415, March 2010.
[2] S. Caorsi, A. Massa, and M. Pastorino, "A crack identification microwave procedure based on a genetic algorithm for nondestructive testing," IEEE Transactions on Antennas and Propagation, vol. 49, no. 12, pp. 1812–1820, Dec. 2001.
[3] G. Wang, C. Gu, T. Inoue, and C. Li, "A Hybrid FMCW-Interferometry Radar for Indoor Precise Positioning and Versatile Life Activity Monitoring," IEEE Transactions on Microwave Theory and Techniques, vol. 62, no. 11, pp. 2812-2822, Nov. 2014.
[4] A.Fedeli, M. Pastorino, C. Ponti, A. Randazzo and G. Schettini, "Forward and Inverse Scattering Models Applied to Through-Wall Imaging," 2020 14th European Conference on Antennas and Propagation (EuCAP), Copenhagen, Denmark, pp. 1-4, 2020
[5] C.Estatico, A. Fedeli, M. Pastorino and A. Randazzo, "Microwave Imaging in Stratified Media: A Multifrequency Inverse-Scattering Approach," 2019 13th European Conference on Antennas and Propagation (EuCAP), Krakow, Poland, pp. 1-4, 2019
[6] X Chen, Computational Methods for Electromagnetic Inverse Scattering. Hoboken, NJ, USA: Wiley, 2018
[7] R. Kleinman and P. Van den Berg. "A modified gradient method for two-dimensional problems in tomography," Journal of Computational and Applied Mathematics, vol.42, no. 1, pp. 17-35, Sep 1992.
[8] W. Chew and Y. Wang, "Reconstruction of two-dimensional permittivity distribution using the distorted Born iterative method," IEEE Trans. Med. Imaging, vol. 9, no. 2, pp. 218-225, June 1990.
[9] A. Abubakar, W Hu, P. van den Berg, and T. Habashy, "A finite-difference contrast source inversion method," Inverse Problems, vol.24, no. 6, 065004, sept 2008.
[10] X. Chen, "Subspace-based optimization method for solving inverse-scattering problems," IEEE Transactions on Geoscience and Remote Sensing, vol. 48, no. 1, pp. 42-49, Jan. 2010.
[11] T. Yin, Z. Wei and X. Chen, "Non-Iterative Methods Based on Singular Value Decomposition for Inverse Scattering Problems," IEEE Transactions on Antennas and Propagation, vol. 68, no. 6, pp. 4764-4773, June 2020.
[12] T Lu, K Agarwal, Y Zhong, et al. "Through-wall imaging: Application of subspace-based optimization method," Progress in Electromagnetics Research, vol.102. pp. 351-366, 2010.
[13] X Chen. "Subspace-based optimization method for inverse scattering problems with an inhomogeneous background medium," Inverse Problems, vol.26, no.7, 074007, May 2008.
[14] X. Ye, R. Song, K. Agarwal, and X. Chen, "Electromagnetic imaging of separable obstacle problem," Opt. Express, vol. 20, no. 3, pp. 2206–2219, Jan. 2012.
[15] Z. Wei and X. Chen, "Deep-learning schemes for full-wave nonlinear inverse scattering problems," IEEE Transactions on Geoscience and Remote Sensing, vol. 57, no. 4, pp. 1849–1860, Apr. 2019.
[16] L. Li, G. Wang, F. Teixeira, C. Liu, A. Nehorai, and T. J. Cui, "DeepNIS: Deep neural network for nonlinear electromagnetic inverse scattering," IEEE Transactions on Antennas and Propagation, vol. 67, no. 3, pp. 1819–1825, Mar. 2019.
[17] R. Guo, X. Song, M. Li, F. Yang, S. Xu and A. Abubakar, "Supervised Descent Learning Technique for 2-D Microwave Imaging," IEEE Transactions on Antennas and Propagation, vol. 67, no. 5, pp. 3550-3554, May 2019
[18] Z. Wei and X. Chen, "Physics-inspired convolutional neural network for solving full-wave inverse scattering problems," IEEE Transactions on Antennas and Propagation, vol. 67, no. 9, pp. 6138–6148, Sep. 2019.
[19] X. Ye, N. Du, D. Yang, X. Jin, R. Song, S. Sheng, and D. Fang, "Application of Generative Adversarial Network-Based Inversion Algorithm in Imaging 2-D Lossy Biaxial Anisotropic Scatterer," IEEE Transactions on Antennas and Propagation, vol. 70, no. 9, pp. 8262-8275, Sep. 2022.
[20] X. Ye, Y. Bai, R. Song, K. Xu and J. An, "An Inhomogeneous Background Imaging Method Based on Generative Adversarial Network," IEEE Transactions on Microwave Theory and Techniques, vol. 68, no. 11, pp. 4684-4693, Nov. 2020.
[21] Z. Liu, Y. Lin, Y. Cao, H. Hu, Y. Wei, Z. Zhang, S. Lin, and B. Guo, "Swin Transformer: Hierarchical Vison Transformer using Shifted Windows," Proceedings of the IEEE/CVF International Conference on Computer Vision (ICCV), pp. 10012-10022, 2021.
[22] J. Liang, J. Cao, G. Sun, K. Zhang, L. Gool, and R. Timofte, "SwinIR: Image Restoration Using Swin Transformer," Proceedings of the IEEE/CVF International Conference on Computer Vision (ICCV), pp. 1833-1844, 2021.
[23] Z Wang, X Cun, J Bao, and J Liu, "Uformer: A general u-shaped transformer for image restoration," Proceedings of the IEEE/CVF Conference on Computer Vision and Pattern Recognition (CVPR), pp. 17683-17693, 2022.
[24] X. Ye and X. Chen, "Subspace-based distorted-born iterative method for solving inverse scattering problems," IEEE Transactions on Antennas and Propagation, vol. 65, no. 12, pp. 7224–7232, Dec. 2017.
[25] X Wang, K Yu, S Wu, J Gu, Y Liu, C Dong, Y Qiao, and C Change, "Esrgan: Enhanced super-resolution generative adversarial networks," European Conference on Computer Vision Workshops, pp. 701–710, 2018.
[26] X Wang, L Xie, C Dong, and Y Shan, "Real-esrgan: Training real-world blind super-resolution with pure synthetic data," Proceedings of the IEEE/CVF International Conference on Computer Vision (ICCV), pp. 1905-1914, 2021.
[27] Y Huang, R Song, K Xu, X Ye, C Li, and X Chen, "Deep Learning Based Inverse Scattering with Structural Similarity Loss Functions," IEEE Sensors Journal, vol. 21, no. 4, pp: 4900-4907, Feb. 2021.
[28] A Vaswani, N Shazeer, N Parmar, et al. "Attention Is All You Need," Advances in neural information processing systems, arXiv.1706.03762, 2017.
[29] H Hu, Z Zhang, Z Xie, and S Lin, "Local relation networks for image recognition," In Proceedings of the IEEE/CVF International Conference on Computer Vision (ICCV), pp. 3464–3473, 2019.
[30] G Elsayed, P Ramachandran, J Shlens, and S Kornblith, "Revisiting spatial invariance with low-rank local connectivity," In International Conference on Machine Learning, pages 2868–2879, 2020.
[31] M. Belge, M. E. Kilmer, and E. L. Miller, "Efficient determination of multiple regularization parameters in a generalized L-curve framework," Inverse Problem, vol. 18, no. 4, pp. 1161, 2002.
[32] Q. Meng, D. Ye, J. Huangfu, C. Li, and L. Ran, "Experimental investigation on through-wall imaging based on non-linear inversions," Electron. Lett., vol. 52, no. 23, pp. 1933–1935, Nov. 2016.